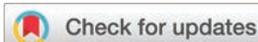

# Rotational polarization nanotopologies in BaTiO₃/SrTiO₃ superlattices†

Saúl Estandía, [a] Florencio Sánchez, [*a] Matthew F. Chisholm [b] and Jaume Gázquez [a]

Ferroelectrics are characterized by domain structures as are other ferroics. At the nanoscale though, ferroelectrics may exhibit non-trivial or exotic polarization configurations under proper electrostatic and elastic conditions. These polar states may possess emerging properties not present in the bulk compounds and are promising for technological applications. Here, the observation of rotational polarization topologies at the nanoscale by means of aberration-corrected scanning transmission electron microscopy is reported in BaTiO₃/SrTiO₃ superlattices grown on cubic SrTiO₃(001). The transition from a highly homogeneous polarization state to the formation of rotational nanodomains has been achieved by controlling the superlattice period while maintaining compressive clamping of the superlattice to the cubic SrTiO₃ substrate. The nanodomains revealed in BaTiO₃ prove that its nominal tetragonal structure also allows rotational polar textures.

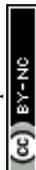

## 1. Introduction

Nanoscale ferroelectrics are different from bulk ferroelectrics mainly because strain and electrostatics such as depolarizing fields play a much more relevant role, but even in the absence of depolarizing fields the long range Coulombian interaction is modified and thickness dependences appear.[1–3] Depolarizing fields, created by unscreened charges at the interfaces, put the system in a non-equilibrium state, and are one of the main reasons of polarization instability.[2,4] Thus, at the nanoscale and under appropriate strain and electrostatic conditions, polar configurations beyond the classical ones may be attained.[5–8] In this regard, the balance of competing energies (elastic, electric, polarization gradient, *etc.*) determines the polarization ground state of the system,[9] and the possibility of fabricating oxide heterostructures with atomic precision has pushed the research into the limits of ferroelectricity at the nanoscale, revealing new phenomena.[10–12] For instance, configurations possessing an emerging toroidal moment were found to exist in ferroelectrics and were proposed as candidates for its use as an alternative way of storing information in ferroelectrics.[7,13] Recently, the experimental observation of long range ordered vortices in PbTiO₃/SrTiO₃ superlattices (PTO/STO SPLs) was accomplished using Scanning Transmission Electron Microscopy (STEM).[14] These vortex topologies present in PTO/STO SPLs are now proving to be a rich system in which properties like chirality,[15] flexoelectricity,[16] or negative capacitance can also be found.[17]

While most experimental works have addressed PTO/STO SPLs or other PTO and BiFeO₃ (BFO) heterostructures,[18–22] studies on exotic states in BaTiO₃ (BTO), one of the most paradigmatic ferroelectrics, are scarce and those that have been done are almost exclusively theoretical investigations.[1] The few recent experimental investigations on BTO have either targeted more conventional domains,[23,24] or lack high spatial resolution.[25] The scarcity of these kinds of studies might be in part due to the fact that a correct determination of the polar distortions in BTO can be more challenging than in PTO, since ion shifts in bulk materials are smaller in BTO ($\delta_{Ti}$ = 9.04 pm, $\delta_{Oequatorial}$ = 4.24 pm, $\delta_{Oapical}$ = 9.85 pm) than in PTO ($\delta_{Ti}$ = 16.02 pm, $\delta_{Oequatorial}$ = 47.3 pm, $\delta_{Oapical}$ = 48.6 pm),[26] which is related to the different ferroelectric modes present in both materials.[27] Nevertheless, new insights were achieved here in BTO grown in the form of $M \times$ (BTO)$_n$/(STO)$_n$ SPLs (where $M$ is the number of BTO/STO bilayers in the SPL, and $n$ is the number of unit cells in each BTO or STO layer). The considered SPLs have $n$ = (2, 4, 10) and $M$ = (30, 15, 6), thus maintaining the total amount of unit cells ($n·M$ = 60). Details regarding the samples' growth and ordinary structural characterization can be found in ref. 28. The SPLs were grown on a STO substrate buffered with a 10 nm thick La₀.₆₇Sr₀.₃₃MnO₃ (LSMO) electrode, which causes a set compressive stress that increases the tetragonality of BTO in its out-of-plane direction, and expect-edly also fixes the spontaneous polarization in this direction. Moreover, when such a ferroelectric is sandwiched between paraelectric layers the discontinuity in the out-of-plane component of the polarization will create large depolarizing fields at the interfaces. How the system responds to such extreme electrostatic boundary conditions depends, *a priori*, on the thickness of the paraelectric layer. For thin paraelectric layers, the depolarizing field can be purged by adopting a uniform polarization across the BTO and the paraelectric layers (either +z or −z). For thicker paraelectric layers, the higher cost of polarizing the paraelectric would induce the formation of a domain structure inside the ferroelectric. For BTO, such domains are anticipated to be alternatively aligned along the +z and −z directions. This model is very similar to the one described by Kittel or Landau–Lifshitz for domains in ferromagnetic systems,[29,30] so they are usually referred to as Kittel domains. Kittel's law, that predicts decreasing domain sizes for decreasing layer thicknesses, also holds true in ferroelectric films,[31] which could make domains below a certain thickness energetically unfeasible. In the same vein, in a BTO/STO SPL with thin paraelectric layers the BTO and STO layers are expected to be electrostatically coupled, as was experimentally demonstrated in BTO/STO SPLs on STO;[32] whereas in a SPL with thicker paraelectric layers the polarization is expected to remain confined within BTO layers. Moreover, chronology may be a decisive factor determining the polar state in a ferroelectric; when the ferroelectric is subject to varying boundary conditions it can get stuck in a polarization state that, as conditions evolve, no longer has the lowest energy. Indeed, it has been shown that the depolarizing fields present during crystal growth may drive the polarization into a multidomain state that remains after screening charges have become available at the end of the crystal growth process.[3,32] The BTO/STO SPLs of different periods grown on STO used in this work showed a continuous reduction in the achievable remnant polarization for increasing periods,[28] which was interpreted in accordance with the model summarized above.

[a] *Institut de Ciència de Materials de Barcelona (ICMAB-CSIC), Campus UAB, Bellaterra 08193, Barcelona, Spain. E-mail: fsanchez@icmab.es*
[b] *Center for Nanophase Materials Sciences, Oak Ridge National Laboratory, Tennessee 37831-6064, USA*

†Electronic supplementary information (ESI) available: Additional figures. See DOI: 10.1039/c9nr08050c

Here, Cs-corrected STEM was employed to go further on the characterization of these SPLs at the unit cell scale. This investigation reveals the transition from a single oriented polarization configuration with expected out-of-plane polarization for short period SPLs ($n$ = 2, 4) to a state that contains rotational nanotopologies for the longest period SPL ($n$ = 10). This occurs in spite of having in-plane compressed SPLs, which is expected to fix the polar axis direction. Moreover, since the STO substrate is cubic, strained BTO/STO SPLs should present 4-fold in-plane symmetry. Therefore, rotational topologies of polarization are not expected *a priori* in these SPLs. However, our results show their formation under suitable SPL periodicity, demonstrating that the feasibility of forming exotic polarization topologies goes beyond heterostructures based on PTO or BFO and could be regarded as a more generalizable behaviour in ferroelectric oxides. Therefore, emerging functional properties as those that have been revealed in PTO-based heterostructures could be pursued in BTO and other ferroelectric oxides.

## 2. Results and discussion

In the following, the STEM characterization of 4 × 4 ((BTO)$_4$/(STO)$_4$) and 10 × 10 ((BTO)$_{10}$/(STO)$_{10}$) SPLs is presented. Results regarding the 2 × 2 SPL can be found in Fig. S1.1 and S1.2, ESI.† Both short-period SPLs (2 × 2 and 4 × 4) have similar domain and structural features. Fig. 1(a) shows a High Angle Annular Dark Field (HAADF) image of the 4 × 4 SPL, where BTO and STO layers can be identified based on the different intensity of Ba and Sr atomic columns, as HAADF imaging mode produces images with a contrast that scales approximately with the square of the atomic number $Z$.[33] In order to identify the ferroelectric domains in the SPLs the atomic displacements of the Ti columns ($\delta_{Ti}$) from the centre of each unit cell were measured, which was determined using the Ba and Sr cation positions. Further details can be found in Fig. S2–S4, ESI.† The resulting dipole map ($\delta_{Ti}$ map) is presented in Fig. 1(b) and reveals that both BTO and STO layers show a continuous polarization with the vast majority of dipoles (yellow arrows) oriented in the out-of-plane (oop) direction. The analysis also provided the lattice parameters of both BTO and STO across the thickness of the SPL. Fig. 1(c) shows the laterally averaged in-plane (ip) and oop lattice parameters, referred to as $a$ and $c$ in the image, respectively. Remarkably, the non-zero $\delta_{Ti}$ in STO indicates that it becomes polarized as well and that STO unit cells may present a small tetragonality (defined as the $c/a$ ratio), see Fig. 1(b) and (d). Tetragonality in STO has been predicted with $c/a$ = 1.008,[34,35] which entails a difference of only 3 pm with respect to the cubic cell, smaller than the experimental error bars. The tetragonality shows a modulation as a result of the variations only in the oop lattice parameter, that alternates between a minimum $c \sim 3.9$ Å in STO and a maximum $c \sim 4.1$ Å in BTO. The ip lattice parameter $a$ does not vary along the thickness and is equal to the STO substrate value $a$ = 3.905 Å (see also Fig. S5, ESI†), indicating that no plastic relaxation takes place. Therefore, STO is almost cubic ($c_{STO}/a_{STO} \simeq 1$) while BTO is tetragonal with a maximum tetragonality of $c_{BTO}/a_{BTO} \sim 1.05$ at the center of the BTO layers. Additionally, Fig. 1(c) reveals a smooth tetragonality gradient that starts at the interface and extends into the BTO layers, while in STO this gradient is restricted to unit cells right at the interface. The smooth tetragonality gradient also entails a unit-cell volume gradient inside BTO, since BTO unit-cells near the interfaces have a smaller oop lattice parameter than those in the centre of the BTO layers. This smoothness is probably driven by an electrostrictive coupling and could favour a smooth polarization variation through the dissimilar layers, thus minimizing the energy associated with polarization gradients.[36] It is worth noting that Electron Energy-Loss Spectroscopy (EELS) atomic-resolution chemical maps (presented in Fig. S6, ESI†) show that the BTO–STO interfaces contain atomic steps of one unit-cell at maximum, thus eliminating the potential effect of averaging atomic steps along the imaging projection direction. Similar smooth strain gradients have been observed in SPLs of PbZr$_{0.2}$Ti$_{0.8}$O$_3$/SrRuO$_3$ and PTO/STO,[37,38] but had not been reported before for BTO/STO SPLs.

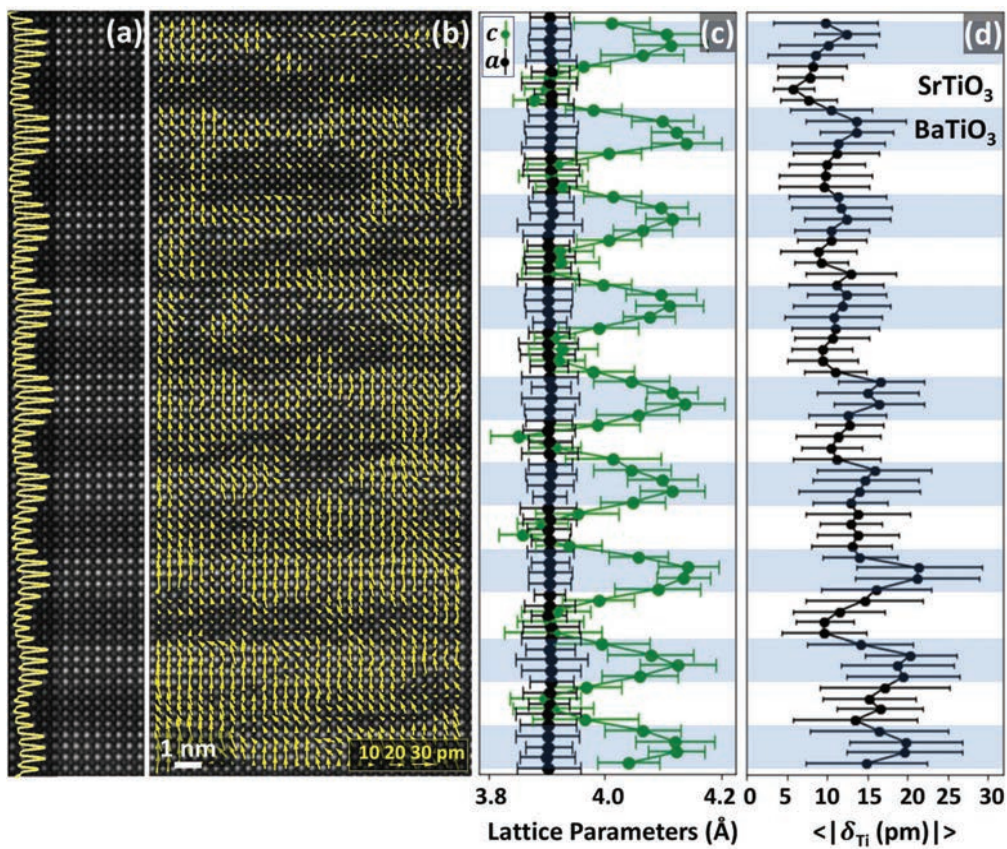

Fig. 1 (a) Slice of the HAADF of the 4 × 4 SPL viewed along the [100] STO zone-axis, with its intensity profile along the oop direction superimposed. (b) Dipoles map superimposed on the HAADF image. (c) Laterally averaged profiles of in-plane and out-of-plane lattice parameters. (d) Laterally averaged profile of the modulus of $\delta_{Ti}$, defined as the distance of the Ti atomic column from the centre of the unit cell. For the sake of clarity, in (c) and (d) the BTO and STO slabs are indicated by coloured stripes. Error bars in (c) and (d) indicate the dispersion of values along each row (see section S15, ESI†).

The laterally averaged $\delta_{Ti}$ modulus, presented in Fig. 1(d), also shows modulations across the SPL. In fact, a comparison of the profiles for the oop lattice parameter (equivalent to $c/a$ shape since $a$ is constant) and $\delta_{Ti}$ shows that they are correlated, as both the peaks and valleys coincide. On top of that, looking at the depth-wise changes of the $\delta_{Ti}$ modulus, Fig. 1(d), one notices that BTO layers closer to the LSMO electrode present much larger $\delta_{Ti}$ values (∼20 pm), and that $\delta_{Ti}$ gradually decreases with increasing distance from the LSMO electrode, to around 10 pm. This effect can be attributed to the metallic character of LSMO, which would provide better screening of polarization charges, with a finite screening length though. In contrast, the less effective screening provided by the bare surface (upper region of the image) and the presence of the paraelectric STO layers induce a sharper reduction of the polarization. Strikingly, the full dipole map in Fig. 1(b) also shows the presence of strong local variations of the imaged $\delta_{Ti}$ modulus, with regions in which $\delta_{Ti}$ is close to zero, which can be due to a change in the polarization modulus or to a change in the direction of polarization (which would result in a smaller projection on the image's plane). These local changes may be originated by the presence of local inhomogeneities such as steps at the interface, intermixing and crystal defects.

While the dipole maps of short period SPLs ($n$ = 2, 4) show that they are in a single oriented polarization configuration, a more complex scenario is found in the 10 × 10 SPL. A full cross-sectional view of the 10 × 10 SPL and its corresponding $\delta_{Ti}$ map are shown in Fig. 2(a) and (b), respectively. Three different regions with distinctive dipole configurations can be distinguished in the $\delta_{Ti}$ map. Those regions at the bottom and at the top of the SPL show most of the dipoles pointing along the oop direction, in particular pointing towards the LSMO at the bottom and towards the vacuum at the top. Within the central region, comprising three BTO layers, the dipole map shows complex topologies of electrical polarization. However, the averaged profiles of oop and ip lattice parameters (Fig. 2(c)) show a modulation of the laterally averaged oop lattice parameter $c$ with strong local variations, just like in the 4 × 4 SPL, although the analysis reveals broader regions in BTO with reduced tetragonality and polarization, extending over 3–5 unit cells into the layers. Moreover, the maximum differences between BTO and STO $c$ parameters remain similar along the thickness of the SPL. EELS analysis, like in shorter

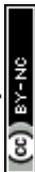

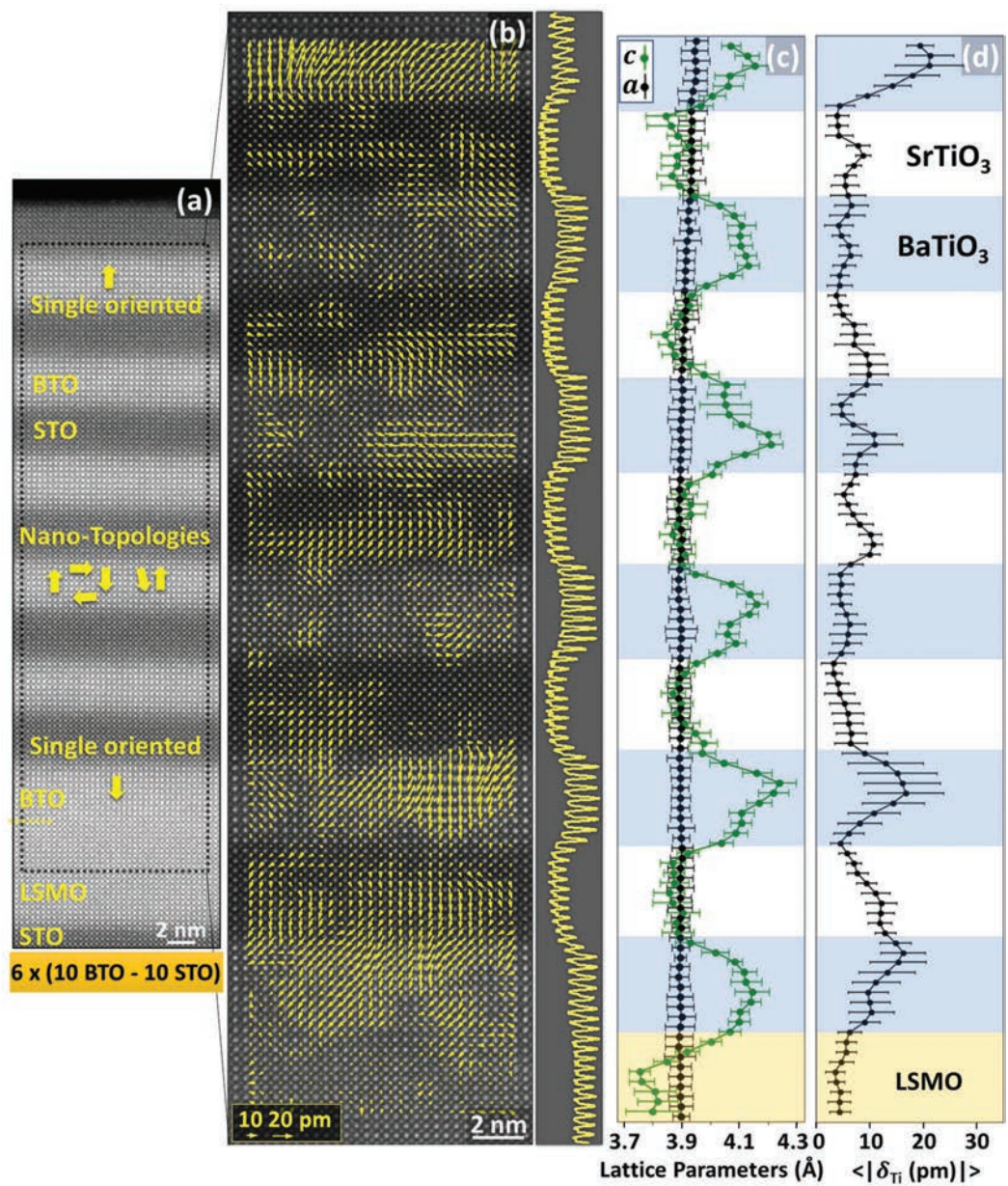

Fig. 2 (a) High field of view HAADF image of the 10 × 10 SPL along the [100] STO zone-axis with superimposed schematics describing the polarization directions. The polar displacement map for the marked area in (a) is shown in (b), together with the HAADF image and its intensity profile (right panel). (c and d) Show the averaged profiles along the SPL thickness of oop and in-plane lattice parameters and of the modulus of $\delta_{Ti}$, respectively. The BTO, STO and LSMO locations are indicated approximately by coloured stripes. Error bars in (c) and (d) indicate the dispersion of values along each row (see section S15, ESI†).

period SPLs, also shows that roughness or interdiffusion is confined to within ±1 unit cell of the BTO/STO interface and cannot account for the volume gradient observed in the 10 × 10 SPLs (see Fig. S6, ESI†).

A careful look at the $\delta_{Ti}$ map, Fig. 2(b), reveals that STO layers present regions with a non-zero oop polarization, in particular at the bottom of each STO layer, while the upper STO/BTO interface shows a lower polarization. The average $\delta_{Ti}$ measured in STO is mostly restricted to the oop direction, and it is rather small (5–9 pm) and only slightly above the noise level (3–4 pm) (see Fig. S7.1 and S7.2, ESI†). The averaged profile of the modulus of $\delta_{Ti}$, Fig. 2(d), also shows a gradient in the polarization across all STO layers, which might be related to two factors. First, that the lower and the upper interface with the ferroelectric have different polarization charges (EELS elemental maps show a more abrupt lower interface, Fig. S6 ESI†). Secondly, that both interfaces have a different unit cell volume (Fig. 2(c)). The latter can also be analysed using the energy-loss near-edge structure (ELNES) of the Ti L-edge across the BTO/STO interfaces.

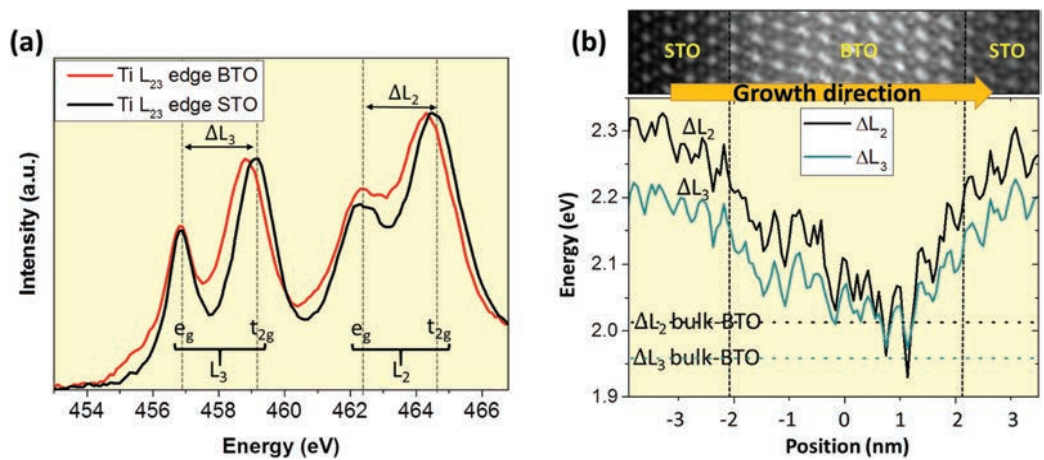

**Fig. 3** (a) Ti L-edge spectra acquired in BTO and STO layers of the 10 × 10 SPL. Due to the oxygen octahedra crystal field, the Ti $L_2$ and $L_3$ peaks are split in energy and allow for obtaining the $e_g$ and $t_{2g}$ 3d sub-bands. The crystal field splitting $\Delta$ for the $L_2$ and $L_3$ are indicated. (b) The upper panel shows the region of the 10 × 10 SPL from which a spectrum image was acquired. The lower panel shows the oop profile of the crystal field splitting $\Delta L_2$ (in black) and $\Delta L_3$ (in green) as a function of position across the BTO and STO layers of the 10 × 10 SPL (upper panel). Changes in the crystal field splitting indicate changes in the Ti–O bonds.

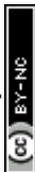

In this direction, Fig. 3(a) shows the Ti L-edge spectra acquired in BTO and STO layers of the 10 × 10 SPL. Notice that the crystal field splitting $\Delta$ for the $L_2$ and $L_3$ are slightly different between BTO and STO. The crystal field strength increases when the distance of Ti–O bonds is reduced, and therefore it is a useful parameter to track the strain state, assuming that a bigger tetragonality means a bigger unit cell volume and therefore a larger Ti–O distance.[38–40] Fig. 3(b) (upper panel) shows a spectrum image across STO and BTO layers. The signal was averaged and the fine structure of the spectra was analysed one atomic plane at the time. Finally, the $\Delta L_2$ and $\Delta L_3$ were calculated, obtaining a smooth variation across STO and BTO layers. This result agrees with the probed smooth tetragonality changes across the interfaces. Notice that the slope is greater in the lower STO/BTO interface, which is where the STO presents a higher polarization (see Fig. 2(b)) and also where the BTO rapidly reaches the $\Delta L_2$ and $\Delta L_3$ values of the bulk. Therefore, dissimilar upper and lower interfaces become evident from both polarization and EELS analysis.

All the observed areas of the 10 × 10 SPL STEM specimen show similar exotic dipole configurations in the SPL central region. A dipole map obtained around the central SPL region is presented in Fig. 4(a), where rotational nanotopologies are visible in BTO (see also Fig. S8, ESI†). An angle colormap is used to assign a different colour to each dipole depending on its direction, which highlights the different polarization directions. The magnitude of $\delta_{Ti}$ in the central area where the rotational topologies are present is between 5 and 10 pm, which is substantially smaller than $\delta_{Ti}$ at the top and bottom of the SPL, where it can reach more than 20 pm, similar to what is found for short period SPLs. This might be due to the fact that, in regions with rotational topologies, the measured $\delta_{Ti}$ is merely a projection. Indeed, the cross-sectional images only have access to two polarization components that lay on the observation plane, which, at the same time, is the resulting projection of several unit cells. Therefore, to attain the correct dipole configuration depends on the domain's dimension and on the imaged STEM-specimen thickness (see S9, ESI†). As visible in Fig. 4(a), different kinds of rotational nanotopologies are present in the dipole map, see a zoom in Fig. 4(b). The closed rotations contain a topological defect, that is, a discontinuity in the order parameter (electrical dipoles in the present case) at the core, and can interact with each other as particle-like objects. Other kinds of dipole configurations, such as waves and mushroom-like configurations are also found (for a more complete description of the different configurations see Fig. S10, ESI†). Fig. 4(c) shows an example of a dipole wave ending in a mushroom-like configuration. Very similar closed rotations, waves and mushroom configurations have been previously reported in PTO/STO SPLs.[18] The observed topological defects bear a topological charge (TC) of either +1 (full clockwise rotation of the dipoles upon a clockwise contour loop around the core), or an opposite TC of −1 (full counter-clockwise rotation upon a clockwise contour loop), and are compatible with vortices for a TC of +1 and antivortices in the case of TC −1. Remarkably, all the observed +1 topological defects are found associated to a −1 topological defect, forming pairs with null net TC, of which an example is shown in Fig. 4(b). The null net TC restricts the coupling range of the pair to dipoles located in its vicinity and allows homogeneous polarization far enough from the TC pair, where the dipoles do not discern the separated TC charges. This is, a null TC pair can exist in a matrix of homogeneous polarization (i.e. surrounded by homogeneous downwards oop polarization). Furthermore, this would make a null TC pair energetically favourable with respect to an isolated topological defect. This bonding of oppositely charged defects is described in the statistical xy model for a 2D lattice of rotors, in which dissociation is pre-

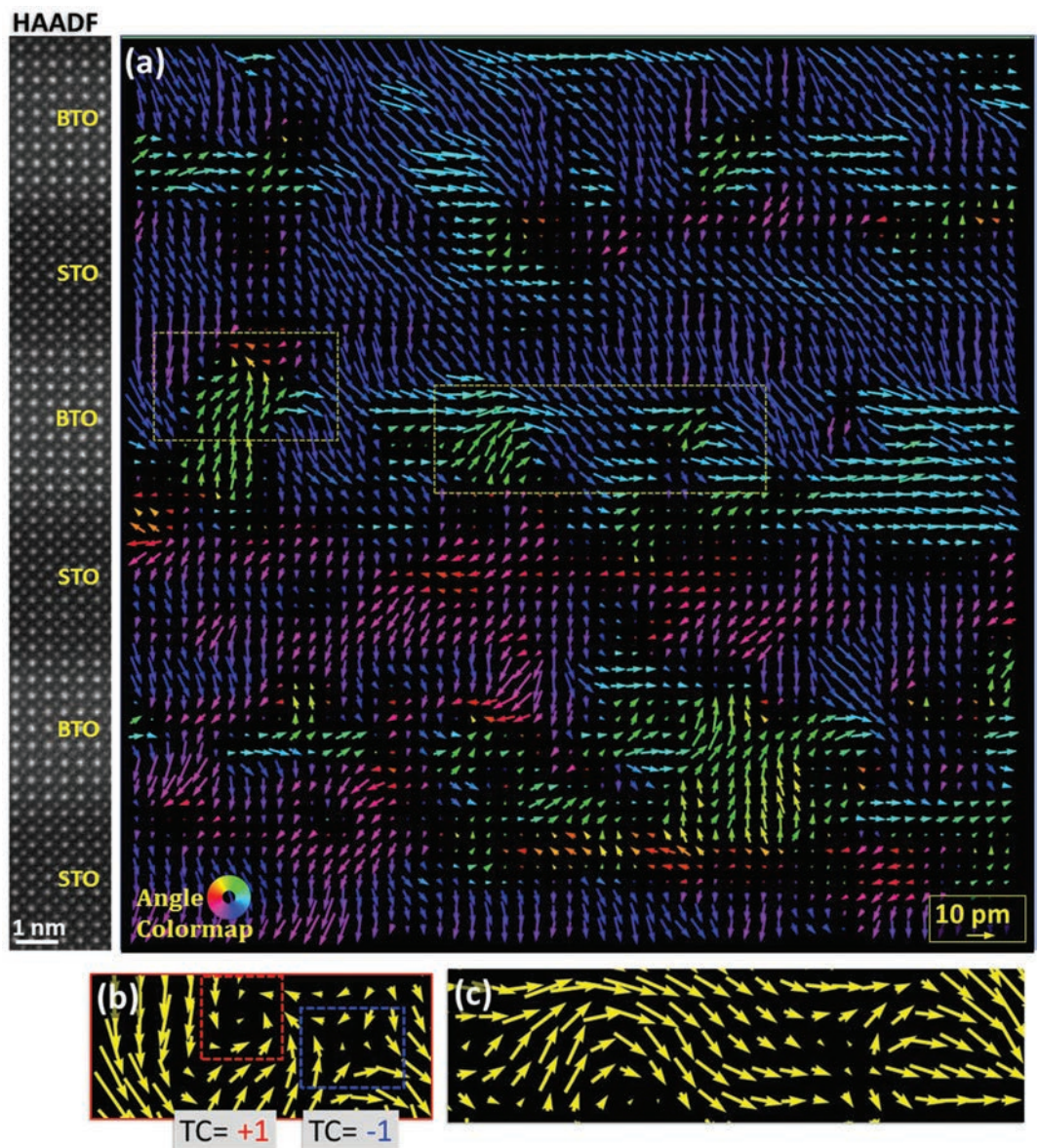

Fig. 4 (a) Dipole maps of an area of the 10 × 10 SPL. As a guide, a slice of the HAADF image from where the dipole map has been extracted is shown at the left of the dipole map. As shown in the legend, the arrows in (a) are coloured according to an angle colour map with the aim of highlighting the different polarization directions. (b and c) Show different zoomed nanotopologies of selected areas marked in (a) (note that angle colormap is not used in (b) and (c)). A pair of topological defects of opposite charge can be appreciated in (b), while (c) shows a dipolar wave ending in a mushroom-like dipole configuration.

dicted to become energetically achievable above the Kosterlitz–Thoules transition temperature,[41] which accounts for the phase transition of different physical systems such as lattices of magnetic spins, liquid crystals, superconductors or superfluids.[42,43]

The transition from a single oriented polarization configuration to an alternating +z and −z domains configuration (both in BTO and STO) was predicted for in-plane compressed BTO/STO SPLs for increasing periods.[44] Instead, our results show that the transition to a polydomain configuration comes in the form of rotational nanotopologies that do not follow the anisotropy of the tetragonal unit cell, suggesting that dipole–dipole interactions dominate over the anisotropy of the crystal and that polarization behaves as a fluid.[45] Alternatively, this could point to the presence of BTO in structural phases other than tetragonal, provided that expectedly BTO unit cell symmetry determines the dipole direction.[24] Other BTO-based ceramics have shown that they can show continuous rotations when there is presence of mixed unit cell symmetries.[46] However, this is not favoured when clamped to the STO substrate with its in-plane 4-fold symmetry. It is remarkable that while the single oriented configuration observed in short period SPLs shows a correlation between the tetragonality and the polarization, no correlation is evident in the areas where rotational nanotopologies are observed (see Fig. S11 and S12, ESI†), obtaining always an oop tetragonality in BTO even when in-plane dipoles are measured. In tetragonal BTO the formation of topologies containing dipoles that considerably deviate from the crystal-axis direction was unexpected from the crystal anisotropy of a BTO/STO SPL clamped to a STO substrate,[38] where a maximum dipole deviation of 45° at the BTO surfaces, where electrostatic conditions make more favourable rotations, was predicted.[47] Therefore, the observed rotational nanotopologies are, *a priori*, not expected to take place in BTO/STO SPLs, which begs the question of the underlying mechanism that permits such a deviation. Predictions of rotational topologies in BTO are restricted to confined stress-free nanostructures such as nanodots and nanowires,[48] while +z and −z domains are predicted for BTO/STO SPLs. Nevertheless, it should be noted that simulations are performed considering mostly perfect BTO crystals, which implies absence of defects, and the existence of small off-stoichiometries cannot be discarded. The present SPLs were grown by pulsed laser deposition, which is a highly energetic technique that can cause the formation of defects.[49] Therefore, the existence of small off-stoichiometries and defects is possible; which could assist the rotation of polarization in BTO, as was found to happen in BFO.[50] However, within the sensitivity of the technique, EELS measurements reveal expected electronic state for titanium and oxygen (see section S6, ESI†). More interestingly, the formation of different topological defects, as for instance bound vortices-antivortices pairs was recently predicted to occur in canonical BTO with different unit cell symmetries (including tetragonal and orthorhombic) at a finite temperature through a new identified mechanism of topological protection.[51] The mechanism allows the stabilization of dipole configurations with dipoles deviating from the crystal anisotropy direction by relying on thermal entropy contributions, and although it was specifically studied in the case of BTO, the result should be generalizable to proper ferroelectrics. Thus, rotational topologies could be expected to be feasible in strained canonical tetragonal BTO.

## 3. Experimental

### Samples

The SPLs were grown by pulsed laser deposition with real-time monitoring by reflection high-energy electron diffraction.[28] The SPLs are $M \times (BTO)_n/(STO)_n$ (where $M$ is the number of BTO–STO bilayers in the SPL, and $n$ is the number of unit cells in each BTO or STO layer). The considered SPLs have $n$ = (2, 4, 10) and $M$ = (30, 15, 6), thus maintaining the total amount of unit cells ($n \cdot M$ = 60). The SPLs were grown on a STO substrate, which has a negative lattice misfit with BTO of −2.23% and thus causes a compressive stress, imposing its in-plane lattice parameter to the SPLs. This lattice mismatch may increase the tetragonality of BTO unit cells and fix its axis along the out-of-plane direction, thus increasing and fixing the spontaneous polarization in this direction. Moreover, it would ensure that no relaxation of BTO to its bulk structure takes place. In addition, the polar discontinuity between BTO (ferroelectric) and STO (paraelectric) puts the system in a high electrostatic energy state that can alter the ferroelectric properties.

### Characterization

The SPLs were characterized by Cs-corrected scanning transmission electron microscopy along the [100] crystallographic axis of the BTO/STO stacks. The samples were prepared for TEM observation by the conventional method of cutting-gluing-slicing-polishing (mechanical + ion milling). STEM images were acquired with an aberration corrected Nion UltraSTEM 200, operated at 200 kV and equipped with a 5th order Nion aberration corrector and with a Gatan Enfinium spectrometer. HAADF and Annular Bright Field (ABF) images were acquired and analysed with self-developed software (see S2 and S13, ESI†). A HAADF image of BTO with a superimposed BTO unit cell is shown in Fig. S2 (ESI†). The high field of view HAADF images of the whole set of SPLs are presented in Fig. S14 (ESI†). Electron Energy-Loss Spectroscopy (EELS) was also used to further characterize the SPLs. Dipole maps, that indicate the polarization direction, were extracted from the titanium shift ($\delta_{Ti}$) measured in the images. For each unit cell, $\delta_{Ti}$ was defined as the difference between the found titanium position and the intersection of the diagonals of barium atomic columns, as schematized in Fig. S2(b).† Also, in-plane and oop lattice parameters ("$a$" and "$c$", respectively) were calculated from the relative position of barium columns. The STEM specimens' thicknesses was measured with EELS, and it was kept in the 0.2–0.3 inelastic mean free paths (25–30 nm) range. In addition, with the illumination conditions used in our work (semi-angle of 30 mrad at 200 kV) the expected depth of field is 55 Å.

## 4. Conclusions

It has been experimentally shown that the paradigmatic ferroelectric, BTO, also exhibits exotic polarization configurations with continuous polarization rotation. The formation of the nanotopologies is achieved through the selection of a suitable SPL period, that controls the electrostatic and mechanical boundary conditions. Remarkably, nanotopologies occur in spite of the BTO high anisotropy. Our results suggest that rotational topologies are the lowest energy configuration in nanosized ferroelectric BTO under certain electrostatic conditions, and may be more feasible than previously thought. These experimental results should encourage further research on exotic polarization topologies in BTO/STO SPLs and other BTO nanostructures.

Finally, some of the emerging properties and potential applications that could be associated to the observed rotational nanotopologies are considered. In a similar manner to vortices in PTO/STO SPLs,[16] some small regions around the topological defects or other topologies could present a stabilized negative capacitance state, since the polar distortion in these locations might become very small, driving the dipoles into the unstable region of negative capacitance that is present in all classical ferroelectrics around the centre of their double-degenerated energy landscape.[52] Not only properties such as chirality could arise,[14] but also other properties could be altered in the regions with rotational nanotopologies. Furthermore, it has been shown that a variety of rotational nanotopologies are stable in BTO at room temperature, and thus there is room to address the question of the feasibility of the control over the different degrees of freedom of the different topologies, which could then point towards similar (or new) technological applications as those of magnetic skyrmions.[53–55]

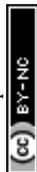

## Conflicts of interest

There are no conflicts to declare.

## Acknowledgements

Financial support from the Spanish Ministry of Economy, Competitiveness and Universities, through the "Severo Ochoa" Programme for Centres of Excellence in R&D (SEV-2015-0496) and the MAT2017-85232-R (AEI/FEDER, EU) project, and from Generalitat de Catalunya (2017 SGR 1377) is acknowledged. The scanning transmission electron microscopy was conducted at Oak Ridge National Laboratory. MFC was supported by the U. S. D. O. E., Office of Science, BES, Materials Sciences and Engineering Division. SE acknowledges the Spanish Ministry of Economy, Competitiveness and Universities for his PhD contract (SEV-2015-0496-16-3) and its cofunding by the ESF. J. G. also acknowledges the Ramon y Cajal program (RYC-2012-11709). SE work has been done as a part of their Ph. D. program in Materials Science at Universitat Autònoma de Barcelona. We thank M. Stengel for useful discussions. We acknowledge support of the publication fee by the CSIC Open Access Publication Support Initiative through its Unit of Information Resources for Research (URICI).

## Notes and references